# Categorising Software Contexts

*Research-in-Progress*

Diana Kirk
School of Computer & Mathematical Sciences
Auckland University of Technology
Private Bag 92006, Auckland 1142, New Zealand
dianakirk@acm.org

Stephen G. MacDonell
Department of Information Science
University of Otago
PO Box 56, Dunedin 9054, New Zealand
stephen.macdonell@otago.ac.nz

**Abstract**

*A growing number of researchers suggest that software process must be tailored to a project's context to achieve maximal performance. Researchers have studied 'context' in an ad-hoc way, with focus on those contextual factors that appear to be of significance. The result is that we have no useful basis upon which to contrast and compare studies. We are currently researching a theoretical basis for software context for the purpose of tailoring and note that a deeper consideration of the meaning of the term 'context' is required before we can proceed. In this paper, we examine the term and present a model based on insights gained from our initial categorisation of contextual factors from the literature. We test our understanding by analysing a further six documents. Our contribution thus far is a model that we believe will support a theoretical operationalisation of software context for the purpose of process tailoring.*

## 1. INTRODUCTION

Ineffective development processes have been identified as contributing towards Information Systems failure. MacCormack et al. suggest that firms must deploy different processes according to business context and that applying a uniform 'best practice' approach results in missed opportunities (MacCormack et al. 2012). Software engineering researchers have indeed found that practitioners frequently adapt and tailor development processes to suit specific project contexts (Avison et al. 2008; Bajec et al. 2007; Fitzgerald 1997; Hansson et al. 2009; MacCormack et al. 2012; Muller et al. 2009; Petersen et al. 2009; Turner et al. 2010). In some cases, tailoring involves adapting a specific methodology, and in others practices are adapted from several approaches, often at the level of the individual project. As an example, as agile approaches have become more established, deficiencies have been exposed, leading to either contextualisation (Hoda et al. 2010) or to amalgamation with other paradigms, for example, the 'lean' paradigm (Wang et al. 2012). The traditional viewpoint that methodologies and practices should be adopted and used as prescribed (Cusumano et al. 2003) has thus been superseded by one of acceptance that tailoring according to project-specific contexts is both necessary and unavoidable.

This state of affairs raises questions about software context. If context is of importance for software practice selection and adaptation, we clearly must strive to fully understand the nature of the relationships between practice and context. Only then will we be in a position to advise industry on which practices might be most suitable or on how to adapt implemented practices.

There have been many efforts thus far to relate project outcomes to key project contextual factors, and several researchers have grouped factors in a bid to model the space in a way that seems to make sense. However, there are an enormous number of possible combinations of factors, and the pragmatic approach taken by researchers means that the resulting models are limited in some fundamental way. In some cases the model is based on inputs from a single organisation and so is inevitably scoped to the operating space of the organisation. The result is that some important contexts are omitted. For example, Avison and Pries-Heje apply an eight dimensional model which includes key ideas such as 'culture', but omits other key contexts, such as 'temporal distance' (Avison et al. 2008). A second kind of limitation occurs when large numbers of factors are categorized along a number of dimensions, but there is a lack of *meaning* in how the categorization occurs. For example, Clarke and O'Connor propose a reference framework containing 44 factors classified along eight dimensions, but the classification 'Cohesion' represents different kinds of meaning, including 'team members who have not worked for you', 'ability to work with uncertain objectives' and 'team geographically distant' (Clarke et al. 2012). A third limitation relates to the scope addressed by the proposed model. Some models are scoped to the *development* project i.e. with implications of spanning requirements determination through to product delivery (Avison et al. 2008; Kruchten 2011). We suggest this scope is not sufficiently broad. The most important practices recommended for new product development involve product determination i.e. occur well before a product development project is commenced. In the 'Software-as-a-Service (SaaS)' delivery paradigm, the emphasis changes from 'developer-driven' to 'customer-driven', where the on-going relationship between development group and customer becomes key (Stuckenberg et al. 2010).

Our viewpoint is that we must take a more theoretical approach if we are to progress. We have created a framework which we are testing for suitability as a theoretical construct for software context (Kirk et al. 2014). Testing takes the form of classifying contextual factors from the literature – if each found factor can be categorised into one-and-only-one framework dimension, we can have some confidence that the dimensions represent a minimal spanning set for the space of software context and would thus be in a position to propose the framework as an operationalisation for context. However, a pilot study exposed many factors that we could not classify. On deeper study of these factors, we understood that one problem was in the definition and scoping of the term 'context'. In this paper, we present our resulting consideration of the term 'context' and test our understanding by analysing contextual factors from six studies from the literature. We believe that this represents a necessary step towards understanding the relationship between software context and practice.

## 2. BACKGROUND

Our overall goal is to provide a decision support mechanism based on our understanding of objectives and contexts that will support the software practitioner in making decisions regarding practice selection. Before this goal can be achieved, we must understand the problem space more deeply. We view software process as a set of practices aimed at meeting specific objectives and that occur within a specific context. Objectives may include anything of relevance to the project, for example, correct functionality, quality, developer satisfaction. When a practice is implemented, the distance to each of the stated objectives is changed. An effective implementation will decrease the distance to all objectives by a maximal amount. An ineffective practice will effect a smaller change, or may even increase distance to some objectives (Kirk et al. 2009).

Our current research addresses context i.e. the environmental factors that affect practice effectiveness. Our aim is to establish an operationalisation for context that will provide a theoretical basis for empirical studies relating to situated practice. As a first step towards achieving this aim, we have created a framework comprising six dimensions that we initially proposed as a suitable construct for context.

The framework structure represents a conceptualization based on two ideas from prior work. The first is an exploration of the boundaries that served to shape and challenge product development in a globally- dispersed, multi-national product development organization (Orlikowski 2002). Our interest in this study lies in its problem-space perspective, exposing boundaries that are temporal (multiple time zones), geographical (multiple global locations), social (many participants), cultural (multiple nationalities), historical (multiple product versions), technical (complex system, variety of standards) and political (different interests). The second idea involves the application of the terms Who, Where, What, When, How and Why when attempting to partition a problem into orthogonal dimensions. Dybå et al. apply this partitioning to understand context, with meanings assigned from organizational science. Zachman created a two-dimensional framework for describing enterprise architecture, using the terms as the first dimension and perspectives on the organization as second dimension (Zachman 2009).

| Dimension | Examples |
|---|---|
| Who | Consistency in world views: affected by nationalities, culture, team structure, power structures, etc. |
| Where | Physical distance; temporal, locational. |
| What | Product-related constraints: affected by standards, external product interfaces, required quality, etc. |
| When | Life-cycle stage of the situated product. |
| How | Engagement constraints: affected by client delivery expectations, expected involvement, etc. |
| Why | Organisational drivers: result in strategies that cause constraints in the other 5 dimensions. |

Figure 1. Overview of framework dimensions

An integration of the above resulted in the framework illustrated in Figure 1.

The meanings of the dimensions have their basis in Orlikowski's work. Our goal was to show that these dimensions represent a minimal, spanning set for the space of software contexts by classifying contextual factors mentioned in the literature (Kirk et al. 2014). The intention was that found factors would expose defects in the framework and would determine lower level structure i.e. sub-dimensions and values for populating these.

Before proceeding with the classification, we clarified what we meant by 'context'. A suitable accepted definition of the term is "The circumstances that form a setting for an event, statement, or idea, and in terms of which it can be fully understood" (Oxford University Press 2014). We adapted this as "The circumstances that form a setting for an organisational initiative and in terms of which outcomes can be fully understood". Based on this definition, we initially defined context as having two aspects:

- Any factor that affects how well a practice meets objectives.
- The factor cannot be changed i.e. represents a hard constraint.

Examples of hard constraints include when specific individuals have been allocated to a project and the project manager can not alter the allocation, and when an organisation has a policy that all acceptance testing will be carried out by a remote, dedicated test team.

Our definition results in the notion that context for one initiative or project may not be context for another. For example, physical location of team members may be fixed in one instance (hard constraint) but may be changeable in another (soft constraint, and so not part of context).

On establishing this definition, we understood that business objectives and implemented techniques and tools are not part of context. The former, although highly important for practice selection, represents a measure of success for a practice (how well did the practice meet the objective) rather than a context that affects how likely it is that the practice will be effective. We view techniques and tools as inherently part of implemented practices (and thus part of the *solution space*) and so these do not represent context. From this reasoning, we note that one of the proposed framework dimensions, *Why*, does not represent a valid dimension because values in this dimension will affect *strategy* which in turn may affect objectives and/or context. This relationship is clear from the figure, as we can see that factors included in this category will affect values in other dimensions i.e. orthogonality is compromised. We consider this more deeply in the next section.

Our pilot study involved sourcing documents from Elsevier's Scopus, applying the search string

*("software development" OR "software engineering" OR "software process" OR "software project" OR "software study" OR "software management") AND (context OR factor) AND (outcome OR success OR failure)).*

This resulted in 2011 documents. We accepted documents that described studies relating to situated process or practice. We rejected studies that:

- did not relate to software organisations or software projects
- focused on process efficacy, IT solutions adoption, products, techniques, metrics or tools
- related to factors for software process improvement
- were not situated in industry.

As our aim was to be as comprehensive as possible, we did not evaluate the studies for quality, but rather accepted any factor that was perceived by the author(s) as affecting outcomes.

After checking abstracts and removing unwanted documents, we were left with 259 documents, from which we analysed three (Kirk et al. 2014). In Figure 2, we show the results for three of the framework dimensions (those with the highest number of factors). We also show, under the section 'Other', those factors that did not clearly map to a framework dimension. Before continuing with the full study, we must understand the reason for the non-compliance of these factors. One possible reason for lack of fit is that the proposed framework is incomplete i.e. does not span the space of software contexts. Another possibility is that we have given insufficient thought to the meaning of the term 'context'. In the next section, we consider this term more deeply.

## 3. RETHINKING CONTEXT

The non-compliance of a significant number of factors led us to realise that we must rethink our understanding of context with a view to explaining the non-compliance. We examined each of the factors and present below the thinking that resulted.

### *Strategy versus Tactics*

As current wisdom implies that software practices must be adapted for individual projects, our desire is to understand context as it relates to a software project or initiative. This means we are interested in the *operational* level rather than the *strategic* one. Organisational strategy involves establishing steps to achieve long-term goals. Steps include definition of objectives, and these are adopted by lower-level management, who decide on tactics to achieve these goals. This established model provides us with some clues as to how to proceed. As the context we are interested in is at the operational level (affecting how well practices meet objectives), we are not concerned in this study with contextual factors that affect strategy, goal-setting and project objectives. We reframe our scope to include only factors relating to tactical considerations and remove strategic factors from our scope.

### *Hard versus Soft Constraints*

In some cases, it is not clear whether a factor represents a hard constraint over which a project has no control or a soft constraint i.e. a local decision has been made by the project manager. For example, the factor 'Functional testing carried out by a separate, remote organisation' may be viewed as a contextual factor (hard constraint) if this represents an organisational policy, but should not be viewed as a contextual factor if the project manager has the power to choose i.e. it is a soft constraint. Indeed, soft constraints actually belong to the *solution* i.e. represent practices that have been selected to meet project objectives. In a 'real' setting, we must take care to include only hard constraints as contexts.

### *Factors that are Multi-Dimensional*

Some of the 'Other' factors in Figure 2 may be considered as being insufficiently precise. For example, we might hypothesise that the factor 'Company size' will not, in itself, affect practice efficacy, but rather must be more accurately stated in terms of cultural and locational aspects. We will consider these as being 'Secondary factors'. We note that each instance of such a factor represents a belief that the factor can, indeed, be fully described by the dimensions of the framework.

### *Factors that are Vague*

Some factors are insufficiently clear as to meaning. For example, the factor 'Requirements certainty' may be the result of one of a number of underlying contexts. Some possibilities are that the client is weak on decision-making ('who'), the client must wait for a remote party before decisions can be made ('where'), and the product to be developed is not clearly understood ('what'). Each of these underlying contexts maps to a different framework dimension. We need to know which one before we can implement a suitable set of practices. For example, creating a prototype to show to the client will have little effect if (s)he is waiting for inputs from a remote party who is on holiday for two weeks (Clear et al. 2012). This category is similar to the previous one, but only one underlying dimension is indicated, we just don't know which one without further investigation.

| Who | Where |
|---|---|
| *Developer* | Company in one country; development in two other countries. |
| Intention to adhere to specific practices. | Cross-functional teams. |
| Understanding of, and expertise in, problem domain. | Distributed dynamic teams. |
| Professional competence. | Insourcing/outsourcing (work done by employees) |
| Technical competence. | On-shore/off-shore (work done in home country of the organisation) |
| Attitude caused by perceptions of age and gender. | Subject matter experts, analysts and project managers at each site. |
| Specialised skills. | Core development outsourced. |
| Individualism. | Different time zones. |
| Uncertainty avoidance. | Shared office hours and office space. |
| Masculinity (assertiveness, competitiveness). | **What** |
| Lack of domain knowledge. | Mobile ICT. |
| Vendor power. | Financial services industry. |
| Business knowledge of vendor. | Application security and privacy issues. |
| Vendor developers' lack of knowledge of the business section. | Telecommunications. |
| Vendor technical skill in IT. | Range of products and services. |
| *Client* | Generic versus custom solutions. |
| Busy schedule. | Payment and expense management solutions. |
| Lack of confidence. | Travel solutions. |
| Lack of motivation. | **Other** |
| Attitude towards relevance of technology. | Strategy. |
| Attitude towards ability to contribute to IT decisions. | Lack of funds. |
| *Relationships* | End users technical infrastructure and working environment. |
| Perception gap between end-users and developers. | Company size. |
| Lack of correlation between analysts' rating of user participation and users' self-rating of involvement. | Open source community. |
| Developing world with abstractive culture. | Modular architecture. |
| Unavailable client. | Modules split between sites (to maximise use of available resources). |
| Lack of common language reference frame. | Customers are individuals through to multi-nationals. |
| Age-gender inequalities. | Roll-out to series of markets / multiple sites in staged approach. |
| Relative positions of parties. | Large-scale, long-term outsourcing. |
| Differences in vocabulary and terminology. | Requirements certainty. |
| Alignment between roles (users, analysts, , developers, testers). | |
| Small close-knit management group. | |

Figure 2. Some factors from the original pilot study.

*Framing for Context*

In overview, our conceptualization of context for the purpose of decision support for software projects involves the following 'rules':

- Only factors that affect operations i.e. implemented practices are included.

- Factors that relate to process, practices and tools, for example, 'agile approach' and 'tool effectiveness', are excluded as we consider these to be a description of the solution, rather than a context that affects solution efficacy.

- Tactical factors that may affect practice efficacy are considered a context only if they
  - are not under the control of the project (hard constraint)
  - do not require more than one framework dimension to describe (not *secondary*)
  - have a clear meaning (not *vague*)

We illustrate the above in Figure 3. Factors relating to strategy, objectives and process are simply out of scope. Tactical factors are candidates only if these represent hard constraints on the initiative. If the candidates are Secondary or Vague, some further analysis is required before mapping onto the framework.

Applying the rules enables us to address all factors in 'Other', apart from 'Modular Architecture'. We must expand the proposed framework dimension 'What' (product characteristics) to account for this factor.

We note that when analysing studies from the literature, it will not always be clear whether a factor is hard or soft. As we are aiming to classify as many factors as possible, we will consider these as hard. We expect that the full study will expose many secondary and vague factors. In the interests of transparency and to support traceability, we will include capture of these in the full study.

As a result of the above analysis, we expand our definition of 'Context' as follows:

- A factor that affects how well a practice meets *tactical* objectives.

- The factor cannot be changed i.e. represents a hard constraint.

- The factor clearly maps to only one framework dimension.

## 4. Pilot to Test Context

In order to confirm our new understanding of the term *context*, we carried out a second pilot based on the already-found documents from Scopus. Of the 259 documents resulting from the preliminary selection, the full texts for 32 were unavailable to us, leaving 227 documents. Our selection strategy involved commencing from the first unselected document (ordered by date and then surname) and selecting each 25th document following. For any document that was unavailable, or did not meet the criteria on reading of the full text, we tried the next document, and so on until we found an acceptable one.

| Strategic | Secondary |
|---|---|
| Strategy. | Inhouse development. |
| Lack of government support. | End users technical infrastructure and working environment. |
| Charging model (fixed price, time and materials, bonuses, penalties). | Company size. |
| Complexity in financial environment of company. | Open source project. |
| Desire to be close to local market. | Customers are individuals through to multi-nationals. |
| Market-driven software company. | Distributed dynamic teams. |
| Product-line approach. | Large-scale, long-term outsourcing. |
| Issues with company-wide strategy. | Project outsourced to third party supplier. |
| Unclear business priority of scope. | Complex, dynamic and inter-connected environment. |
| Funded by virtual-capital firm. | Corporate politics with negative effects on the project. |
| Funded via equity investors. | Organisational boundaries. |
| Funded via IT focused venture fund. | Globally distributed software production |

Figure 4. Context for understanding software practices.

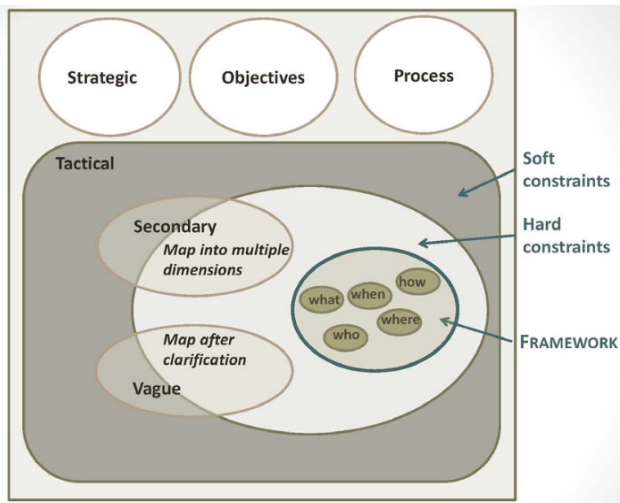

Figure 3. Context for understanding software practices.

The 6 documents analysed are noted in Table 1.

Table 1. Documents analysed in the second pilot study.

| Year | Topic | Reference |
|---|---|---|
| 2013 | Contract model | Atkinson & Benefield |
| 2013 | Project risk analysis (in-press 2011) | Hu et al. |
| 2013 | People in GSD | Misra et al. |
| 2011 | Communications gaps | Bjarnason et al. |
| 2011 | Defect management | Nair et al. |
| 2010 | Off-shore outsourcing success factors | Yalaho & Nahar |

In Figure 4, we show some of the contextual factors classified as *Strategic* and *Secondary*.

In Figure 5, we show some of the factors-of-interest for the framework dimensions *Where* and *What*. Full listings are available on request.

## 5. DISCUSSION

We have earlier created a conceptual framework for software context to support software practice selection. We carried out a trial that exposed several factors that did not clearly fit into the framework. In this paper, we described the thinking that occurred as a result of these categorisation failures. This thinking involved a realization that we had not clearly framed what we meant by 'software context'. The paper represents the outcomes of the reframing activity and the trial carried out to test our new understanding. We were successful in classifying factors that we found in the six documents analysed. We did experience one issue on reading the categorized lists, and this is illustrated in Figures 4 and 5. In Figure 4, we classify 'Globally distributed software development' as Secondary, because the phrase does not provide sufficient information about cultural aspects (who) and specifics about separation (where). In Figure 5, we classify 'Many development teams' as Where, because, in the context of the source document, the inference was that this was a locational issue. However, as a stand-alone term, 'Many development teams' has multiple aspects and is perhaps more suitably classified as Secondary. As we wish to build up comprehensive lists of context factors, the lesson is that we must take care to write down terms in a context-free manner. For example, 'Many development teams' could have been written 'Development teams in different locations', if the meaning relates to locational distance.

The main limitation of our approach thus far concerns the initial selection of documents. We searched only one source, using the search string presented above. Although additional sources are planned for the full study (Kirk et al. 2014), feedback on the work thus far causes us to rethink the scoping of the search string(s) to apply. We believe these must be extended.

In some ways, our study might be considered as the development of a taxonomy for software context. We are reluctant to carry this viewpoint too far because we are not convinced that developing a taxonomy for such a complex and poorly-understood space is an appropriate approach, particularly during the initial exploration stage. However, there are parallels in our approach taken. For example, Nickerson et al. propose a number of steps that must be taken during taxonomy development. These include determining meta-characteristics and ending conditions and then iterating on taxonomy development until ending conditions are met, at each iteration deciding whether the approach will be empirical-to-conceptual or conceptual-to-empirical (Nickerson et al. 2013). Our initial approach was conceptual-to-empirical i.e. we based the framework on ideas from earlier studies and plan to progress based on what we find in the literature. However, although we were clear from the start about the potential users of our framework, we did not understand until we more deeply considered what was meant by context that we needed to scope context to include tactical factors and exclude strategic ones i.e. we were not in a position to establish meta-characteristics until after the first iteration had taken place. We also have an issue with terminology in that many terms in software development are ill-defined and this complicates the task of identifying the characteristics for each dimension. A third issue relates to termination. Once we have found a set of top- level dimensions that do not change, we must still test these by an exhaustive study of factors from the literature, because we are aiming to create a theoretical construct. We submit that these are indicative of the paradigm difference between 'categorising things' and 'categorising how-things-are'. We present these thoughts for general discussion.

| Where | What |
|---|---|
| Location of end users and developers. | *Application Subject Area* |
| Company in one country; development in two other countries. | Mobile ICT. |
| Teams separated according to function. | Financial services industry. |
| Insourcing/outsourcing (work done by employees or not). | Telecommunications.. |
| On-shore/off-shore. | Payment and expense management solutions. |
| Subject experts, business analysts and project managers at each site. | Travel solutions. |
| Core development outsourced. | Business Intelligence (BI). |
| Different time zones. | Video/audio/image processing. |
| Shared office hours and office space. | Healthcare software. |
| Programmers physically separated from designers. | HR and payroll systems. |
| Programmers have no access to designers. | Corporate training administration systems. |
| Programmers have no/little access to customer representative. | Enterprise application integration. |
| Multiple geographic locations. | *Nature of the software* |
| Multi-site. | Application security and privacy issues. |
| Distributed setting. | Range of products and services. |
| People placed close together. | System may interlink with other systems within organisation. |
| People placed around the globe. | System may interface to systems in other organisations. |
| Geographical separation. | Embedded systems. |
| Teams in different countries. | Complex product. |
| Temporal distance. | Diverse technologies. |
| Different development teams. | Size of product. |
| Teams located at different sites. | Generic versus custom solutions. |
| Many development teams. | Multi-tier applications with desktop, web-based and mobile clients. |
| Offshore. | B2B and B2C portals. |
| Various time zones. | *Certainty* |
| Multiple global locations. | Changing technology causes customer to revise thinking. |
| Outsourcing across borders. | System requirements not adequately identified. |
| Outsourcing in same country. | Agreed software specification. |
| | *Structure* |
| | Common code base for product line. |

Figure 5. Some factors-of-interest from the second pilot study.

The key ideas that have emerged from our framing of context are that a) we need to be clear whether we are discussing context-for-strategy or context-for-operations, and b) many factors found in the literature are complex or vague in nature. Our proposed framework relates to context-for-operations, and this understanding will enable us to proceed with our study. However, we have a responsibility to classify all factors found, because a statement that a factor is complex is really a belief based on the assumption that our framework is suitable and a statement that a factor is vague must be justified by giving some possible interpretations.

As our current research is scoped to the operational space (we want to advise on suitability of practices for meeting already-defined objectives), we will not at this stage give much consideration to strategic factors. It is possible that some overlap might occur, in that a contextual factor might affect both strategy and operations. For example, a context 'Insufficient funds' might result in two decisions. The first is to lay off staff in the near future (strategic) and the second is to reduce testing effort in current projects (tactical). This is an area for future research. There is also a possible overlap between *objectives* and the 'What' dimension. For example, 'Quality' describes some aspect of the product to be developed ('What' in our framework) but also may be seen as an objective. It might be that the 'What' dimension needs to be clarified and more thought is required. We hope that the full study will help clarify this.

In summary, our research objective is to support software organisations in practice selection according to local contexts, and we have developed a framework that we propose as a theoretical model for an operationalization of context. An initial test of the framework exposed some issues that caused us to seek a deeper understanding of the term 'Context'. In this paper, we have provided some thoughts on what constitutes 'Context' for the purpose of understanding situated software process. Factors that are out-of- scope include those that relate to objectives, tools or techniques. Factors that are in-scope but cannot be used without further analysis include those that can be split into more than one framework dimension and those that are vague in nature. We have analysed six documents from the literature to test our ideas. As the documents analysed covered a wide range of topics, we feel a measure of confidence in our approach thus far. We plan to implement the full study to test the top-level dimensions of our framework for suitability as a theoretical construct for software context. Future work will involve a consideration of sub-dimensions.